\documentclass[lettersize,journal]{IEEEtran}
\IEEEoverridecommandlockouts
\usepackage{cite}
\usepackage{amsmath,amssymb,amsfonts}
\usepackage{graphicx}
\usepackage{textcomp}
\usepackage{xcolor}
\usepackage{array}
\usepackage{booktabs}
\usepackage{multirow}
\usepackage{caption}
\usepackage{subcaption}
\usepackage{algorithm}
\usepackage{algpseudocode}
\usepackage{hyperref}
\hypersetup{
    colorlinks=true,
    linkcolor=blue,
    filecolor=magenta,      
    urlcolor=cyan,
    citecolor=blue
}
\title{Spreading the Wave: Low-Complexity PAPR Reduction for AFDM and OCDM in 6G Networks}
\author{
\IEEEauthorblockN{Afan Ali, Abdelali Arous,~\IEEEmembership{Graduate~Student~Member,~IEEE},   and H\"{u}seyin Arslan,}
\IEEEmembership{Fellow, IEEE}
\thanks{The authors are with the Department of Electrical and Electronics Engineering, Istanbul Medipol University, Istanbul, 34810, Turkey (e-mail: afanali85@gmail.com; abdelali.arous@std.medipol.edu.tr; huseyinarslan@medipol.edu.tr).}
}

\begin{document}
\maketitle
\begin{abstract}

High Peak-to-Average Power Ratio (PAPR) is still a common issue in multicarrier signal modulation systems such as Orthogonal Chirp Division Multiplexing (OCDM) and Affine Frequency Division Multiplexing (AFDM), which are envisioned to play a central role in 6G networks. To this end, this paper aims to investigate a novel and low-complexity solution towards minimizing the PAPR with the aid of a unified premodulation data spreading paradigm. It analyze four spreading techniques namely, Walsh-Hadamard transform (WHT), Discrete Cosine transform (DCT), Zadoff-Chu transform (ZC), and Interleaved Discrete Fourier transform (IDFT), which assist in preallocating energy prior to OCDM and AFDM modulation. The proposed method takes advantage of the inherent characteristics of chirp-based modulation to achieve a notable reduction in PAPR at minimal computational load and no side information as compared to past solutions, such as Partial Transmit Sequence (PTS) or Selected Mapping (SLM), which suffers with a high computational complexity. The proposed method has an additional benefit of achieving an improvement in phase selectivity by increasing chirp parameters of AFDM and quadratic phase of OCDM, which amplifies the robustness in doubly dispersive channels. It further reduces interference by smoothing the output spread signal. The analytical and simulation results demonstrate an improvement in the overall energy efficiency and scalability of large ioT sensor networks.
\end{abstract}


\section{Introduction}
The rapid emergence of wireless networks has spurred the demand for stable, efficient modulation techniques that can tackle the challenges introduced by doubly dispersive channels. Orthogonal Frequency Division Multiplexing (OFDM) has been a stalwart in standards such as Long-term Evolution (LTE) and Wireless Fidelity (Wi-Fi) due to its efficiency in spectrum usage and its resilience to multipath fading \cite{63}. However, its large Peak-to-Average Power Ratio (PAPR) is a crippling limitation, necessitating power amplifier back-off and energy efficiency trade-off \cite{2,3,5,8}. New paradigms like Orthogonal Chirp Division Multiplexing (OCDM) and Affine Frequency Division Multiplexing (AFDM) have been recently proposed for exploring their ability to utilize chirp-based modulation in order to achieve enhanced diversity and orthogonality in linear time-varying (LTV) channels which are doubly dispersive \cite{6,9,34,36,38,39}. Despite the advantages, OCDM and AFDM also suffer from OFDM-like PAPR issues, which are exacerbated by their complex chirp patterns. However, very less literature is found for PAPR reduction in chirp based waveforms.

PAPR reduction in upcoming waveforms is crucial for future networks, particularly with 6G research focus on high-rate, low-latency communication in dynamic environments \cite{5,8}. Traditional techniques for OFDM, namely Clipping and Filtering, Partial Transmit Sequence (PTS), and Selected Mapping (SLM), have been well researched and optimized \cite{8,11}. In \cite{12}, authors describe a suboptimal approach for PAPR reduction using SLM and PTS with improved PAPR statistics. However, these schemes typically introduce a very high computational complexity, distortion, or side information overhead. More recent research has explored tailored solutions, such as chirp parameter optimisation in AFDM \cite{6} and phase optimisation in OCDM \cite{38,39}, although these remain computationally intensive and are not system-independent. Machine learning (ML)-based techniques have also emerged in recent times, with deep neural networks enhancing PAPR reduction in OFDM by learning signal characteristics~\cite{26,27,29}. In~\cite{7}, authors propose a novel ML-based solution, called PAPRer, which automatically and accurately tune the optimal PAPR target for frequency-selective PAPR reduction. Similarly, authors in \cite{28} suggest a model-driven deep learning algorithm for PAPR reduction in
OFDM by an iterative peak-canceling signal generation scheme. Spreading techniques have been employed in 
Discrete Fourier Transform-spread OFDM (DFT-s-OFDM), which reduce PAPR by spreading data across subcarriers through a DFT precoding step, effectively reducing peak power by single-carrier-like behavior while preserving multicarrier advantages~\cite{15,30,31,32}. Although many more mature spreading techniques exist such as Walsh-Hadamard Transform (WHT), Discrete Cosine Transform (DCT), Zadoff-Chu (ZC) sequences and interleaved DFT (IDFT), most work done to reduce PAPR in OFDM use only DFT as the spreading method, neglecting other spreading methods for this purpose. The authors in \cite{52} design a preamble comprising of a ZC sequence to improve the timing synchronization of OFDM systems, rather than exploring its added effect on PAPR. Similarly, authors in \cite{61} proposes DCT as a signal compression tool. Moreover,, their application to chirp-based waveforms like OCDM and AFDM remains underexplored.

This article investigates new dimension to spreading techniques as a low-complexity PAPR reduction method for OCDM and AFDM. These techniques are used in isolation for various other purposes such as timing synchronization or signal compression but not to explore its effect in PAPR reduction. Consequently, four spreading techniques, mainly, WHT, DCT, ZC sequences, and IDFT, are proposed as a premodulation step to mitigate PAPR effect. As compared to conventional techniques, the proposed spreading method distributes energy of a signal over the entire bandwidth mitigating peaks, without iterative optimization or overhead signaling. Based on established applications of the techniques, WHT in CDMA \cite{59,60}, DCT in signal compression \cite{61}, ZC in LTE synchronization \cite{62}, and IDFT in OFDM \cite{63}, the proposed system adapts them to enhance phase selectivity, reduce complexity, and avoid interference, as validated by mathematical analysis and conceptual illustrations.

The contribution of this work lies in its simplicity and versatility, meeting a gap in PAPR reduction research for chirp-based systems. Although recent studies have expanded the frontier of modulation design for 6G \cite{64,65,66}, few have tackled OCDM and AFDM's PAPR with such an efficient and integrated approach. This paper contributes to the literature with a comprehensive evaluation of the proposed framework, citing state-of-the-art developments, as a promising candidate for next-generation wireless networks with high mobility.

The main contribution of this work can be summarized as follows:
\begin{itemize}
    \item While most PAPR reduction techniques are limited to legacy OFDM waveform and have high computational complexity, we are the first to present detail analysis of PAPR in chirp-based waveforms mainly: AFDM and OCDM, considering their practicality due to robustness in doubly dispersive channel. The study indicates that both AFDM and OCDM also suffers from high PAPR due to their inherent multicarrier nature. Based on that, we propose  a unified premodulation data spreading paradigm that employs WHT, DCT, ZC, and IDFT to pre-allocate energy prior to OCDM and AFDM modulation with remarkable PAPR mitigation.
    \item We demonstrate that our proposed spreading method has a low complexity and requires no side information. On the contrary, conventional techniques such as PTS or SLM incur \(O(N^M)\) complexity or side information, our technique takes advantage of fast algorithms (e.g., \(O(N \log N)\) for WHT, IDFT) and overhead elimination, facilitating its flexibility in chirp-based systems.
    \item As an additional advantage, we show that our proposed framework provides increased phase selectivity and interference for output spread signal. The introduced spreading technique increases the chirp parameters of AFDM (\(c_1, c_2\)) and the quadratic phase of OCDM, thus amplifying the robustness of the doubly dispersive channel. It further reduces interference by smoothing output signal. Techniques like WHT and ZC make use of orthogonality to decrease inter-symbol and inter-carrier interference, enhancing AFDM's diversity and OCDM's chirp separation, representing an energy-efficient solution for high-density high-mobility networks.
    \item Lastly, our work investigates an in-depth analysis of energy efficiency by comparing the proposed OCDM and AFDM across five cumulative distribution functions (CCDF) for PAPR. Therefore, our work presents a novel analysis in terms of energy savings (in MWh) and $CO_2$ (in metric tons/sensor) for emission reduction, in OCDM and AFDM systems. We show that our proposed spreading method make it an energy efficient and scalable for large IoT sensor networks.
\end{itemize}

The paper is organized as follows: In Section \ref{sec:system_model}, the description of the system model is given alongwith available PAPR reduction techniques.  Section \ref{sec:waveform_spread} introduces four spreading techniques and their effect on reducing PAPR and overall system complexity. Then, a novel PAPR reduction framework is proposed in \ref{sec:proposed}, focusing on its additional benefits in terms of improvement in phase selectivity and interference. Section \ref{sec:simulations} provides a performance analysis for the proposed framework. Finally, Section \ref{sec:conclusion} concludes the paper.

\section{System Model}
\label{sec:system_model}
This section outlines the characteristics of AFDM and OCDM waveform, highlighting their inherent tendency for high PAPR due to multicarrier communication system. It further delves into conventional PAPR reduction techniques for legacy OFDM communication system, highlighting little work done for PAPR reduction in OCDM and AFDM.

\subsection{AFDM and OCDM}

Both AFDM and OCDM are multicarrier systems that utilize chirp-based waveforms \cite{6,9,10,19}. Their waveform filters can be expressed in terms of subcarrier index (\(k\)) and time index (\(n\)), over a total of \(N\) samples.

\begin{figure}[t!]
\centerline{\includegraphics[width=1\columnwidth]{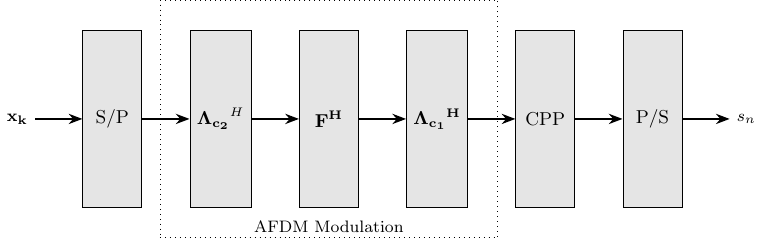}}
    \caption{Block Diagram for AFDM modulation.}
    \label{fig:afdm-BD}
\end{figure}

 In AFDM, each subcarrier is spread on the time-frequency (TF) plane by the inverse discrete affine Fourier transform (IDAFT). Fig.~\ref{fig:afdm-BD} illustrates the block diagram of the AFDM transmitter, showing each transformation block. Chirp-periodic prefix (CPP) is used to deal with multipath propagation \cite{10}.
 
 The waveform filter for AFDM, $g_{k,n}^{AFDM}$ is expressed as:

\begin{equation}
g_{k,n}^{AFDM} = \frac{1}{\sqrt{N}} e^{j2\pi \left( c_1 k^2 + \frac{1}{N} k n + c_2 n^2 \right)}, \quad k, n = 0, 1, \ldots, N-1
\label{eq_afdm}
\end{equation}
where \(c_1\) and \(c_2\) are the chirp parameters that control the quadratic phase terms, \(\frac{1}{\sqrt{N}}\) is applied to normalize.
The input signal \(x_k\) is transformed by waveform filter $g_{k,n}^{AFDM}$ to give output signal \(s_n\) in time-domain as follows:
\begin{equation}
s_n = \sum_{k=0}^{N-1} x_k g_{k,n}^{AFDM},
\label{eq_afdm_transform}
\end{equation}
To achieve total diversity in integer Doppler shift LTV channels, it is required that \(c_2\)  be irrational, and the minimum \(c_1\) is defined as:
\begin{equation}
c_1 = \frac{2\alpha_{\text{max}} + 1}{2N}
\label{c1}
\end{equation}
where \(\alpha_{\text{max}}\) is the integer part of the maximum normalized Doppler shift. The arbitrariness in \(c_2\) can be utilized to reduce PAPR, while \(c_1\) is more constrained. The quadratic terms \(c_1 k^2\) and \(c_2 n^2\) produce chirp behavior, and the linear term \(\frac{1}{N} k n\) is comparable to OFDM's frequency allocation. In matrix notation, DAFT matrix \(\mathbf{\Lambda}\) is expressed as
\begin{equation}
\mathbf{\Lambda} = \mathbf{\Lambda_{c_2} F \Lambda_{c_1}}, \label{eq:daft_matrix}
\end{equation}
where \(\mathbf{F}\) is the \(N \times N\) DFT matrix, and \(\mathbf{\Lambda_{c}}\) is a diagonal matrix defined as
\begin{equation}
\Lambda_{c} = \text{diag}(e^{-j 2\pi c n^2}, \, n = 0, 1, \ldots, N-1), \label{eq:lambda_c}
\end{equation}

\begin{figure}[t!]
    \centerline{\includegraphics[width=1.0\columnwidth]{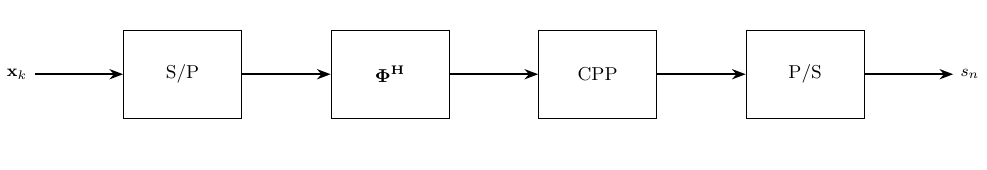}}
    \caption{Block Diagram for OCDM modulation.}
    \label{fig:ocdm-BD}
\end{figure}

The block diagram of OCDM transmitter is depicted in  Fig.~\ref{fig:ocdm-BD}. In OCDM, frequency varies linearly with respect to time, and the phase is quadratic. Assuming \(N\) is even, the waveform filter, $g_{k,n}^{OCDM}$ is:
\begin{equation}
g_{k,n}^{OCDM} = e^{ \left(-j \frac{\pi}{N} (n - k)^2 + j \frac{\pi}{4}\right)}, \quad k, n = 0, 1, \ldots, N-1
\label{eq_ocdm}
\end{equation}
where the term \(-j \frac{\pi}{N} (n - k)^2\) determines the quadratic phase variation, \(j \frac{\pi}{4}\) is a constant phase shift. 
The output signal \(s_n\) is written by inserting waveform filter from \eqref{eq_ocdm} to \eqref{eq_afdm_transform} to get:

\begin{equation}
s_n = \sum_{k=0}^{N-1} x_k g_{k,n}^{OCDM}
\label{eq_ocdm_transform}
\end{equation}

To express \eqref{eq_ocdm} in a matrix-vector form, we define \(\mathbf{\Phi}\) as the \(N \times N\) unitary discrete Fresnel transform (DFnT) matrix. Thus, \eqref{eq_ocdm} can be rewritten as:
\begin{equation}
\mathbf{s} = \mathbf{\Phi^H x_k}, \label{eq:ocdm_matrix}
\end{equation}
where the \((k, n)\)-th element of \( \mathbf{\Phi=g_{k,n}}\) in \eqref{eq_ocdm}.

Both AFDM and OCDM suffer from high PAPR due to their multicarrier nature, where multiple orthogonal subcarriers are summed to form the transmitted signal.  The superposition of \( N \) subcarriers, each with random phase and amplitude, leads to constructive interference, causing large peaks in \( |s_n|^2 \). The PAPR is defined as:

\begin{equation}
 \text{PAPR} = \frac{\max_m |s_n(n)|^2}{\frac{1}{N} \sum_{n=0}^{N-1} |s_n(n)|^2} ,
 \label{papr_eq}
\end{equation}

PAPR can grow as large as \( N \) in the worst case, necessitating PAPR reduction techniques for efficient power amplification.

\subsection{Conventional PAPR reduction techniques}
The problem of high PAPR is well-tackled in literature for OFDM systems~\cite{5,7,8,11}. Some work is recently proposed for PAPR reduction in OCDM and AFDM~\cite{4,38,39,40}. However, each of these techniques suffers from certain limitations.
Some well-known PAPR reduction techniques and their limitations are briefly explained as follows:

\subsubsection{Clipping and Filtering}
     Clipping and Filtering reduces PAPR by limiting \(|s_n|\) to a clipping level (CL) threshold \(\beta\) using a clipper \cite{1,5,14}. If the output signal exceeds  \(\beta\), then the signal is clipped; otherwise, the signal does not suffer any change. 
     The output signal  $s_n'$ in this case can be written as:
    \begin{equation}
    s_n' =
    \begin{cases}
    s_n, & \text{if } |s_n| \leq \beta \\
    A e^{j2\pi arg(s_n)}, & \text{if } |s_n| > \beta
    \end{cases}
    \end{equation}
   Any distortions that arises in $s_n'$ due to clipping is then reduced by using filtering.

\subsubsection{Partial Transmit Sequence (PTS)}: In this technique, data is partitioned into \(M\) sub-blocks, which are independently modulated. After performing inverse Discrete Fourier Transform (IDFT) for each sub-block, output is weighted by a phase factor which is optimized to minimize PAPR to give the following output signal $s_n^{'}$ \cite{12}:
    \begin{equation}
    s_n^{'} = \sum_{m=0}^{M-1} b_m s_{n}, \quad 
    \end{equation}
In this method, optimal phase factor, \(b_m\), minimizes PAPR but requires side information which may increase spectral efficiency.
   
\subsubsection{Selected Mapping (SLM)}:
This technique generates alternative candidate signals, \(u\), from a single source by introducing variations based on different phase sequences. These candidate signals undergoes adaptive optimization techniques,
and then the signal with the minimum PAPR is transmitted. It can be written as follows \cite{25}:
    \begin{equation}
    s_n^{u} = \sum_{k=0}^{N-1} x_k e^{j\phi_k^{(u)}} g_{k,n}
    \end{equation}
    where $\phi_k^{(u)}$ represents phase variation introduced by \(u\).

The signal with the lowest-PAPR signal is transmitted and therefore, it also needs side information.
   
\subsubsection{Tone Reservation (TR)}: 
TR is another technique that focuses on allocating a portion of the available
subcarriers in an OFDM system for the inclusion of additional tones. These reserved
cancellation signals \(c_n\) are designed to nullify high-power peaks in the transmitted signal. Therefore. the new output signal $s_n^{'}$ can be written as \cite{8,11}:
    \begin{equation}
    s_n = s_n + c_n
    \end{equation}
TR helps to reduce the throughput due to non-data tones.

\subsubsection{Machine Learning (ML)}:
ML algorithms are also vastly investigated for PAPR reduction in OFDM systems. It demonstrates remarkable capabilities in finding complex patterns within PAPR levels, making accurate predictions about future PAPR levels~\cite{67,68,69}. It has a ability to dynamically adjust transmission parameters in real time which ensures a continuous optimization process that
minimizes PAPR levels while preserving the optimal signal quality.

\subsubsection{Chirp Selection Technique}
This is a recently proposed technique for OCDM. It is based on selection of frequency variations of the chirps (up or down). Two OCDM signals are generated with up and down chirps, respectively. Then a subset of chirp bases with lower peak contributions is selected, modifying \(g_{k,n}^{OCDM}\) \cite{40}. This adjustment of the quadratic term’s coefficient could be written as follows to get a new waveform filter $g_{k,n}'$:
    \begin{equation}
    g_{k,n}' = e^{(-j \frac{\pi \alpha}{N} (n - k)^2 + j \frac{\pi}{4}})
    \end{equation}
    where \(\alpha\) is optimized. 
However, this method also suffers from high complexity in searching optimal bases to optimize PAPR.

\subsubsection{Clipping with Chirp Optimization}
The is also a relatively new technique for chirp-based waveforms that first optimize the chirp parameters such as, chirp rate, initial phase and amplitude, to lower initial PAPR and then apply clipping to minimize the remaining peaks. This is an iterative method that refines chirp parameters to minimize distortion. Consequently, \(s_n\) is clipped and final output signal  $s_n'$ is written as follows:
    \begin{equation}
    s_n' = P e^{j2\pi\arg(s_n)} + \sum_{k=0}^{N-1} x_k g_{k,n} e^{j2\pi\theta_k}
    \end{equation}
where $P$ and $\theta_k$ are the adjusted amplitude and phase respectively.\\
This technique suffers from increased distortion. Moreover, iterative optimization of chirp parameters can increase computational overhead.
    
\subsubsection{Grouped Pre-Chirp Selection}
This is a recent PAPR reduction method proposed for AFDM systems. In this technique, subcarriers are grouped varying the pre chirp parameter in a non-enumerated manner. Then, the signal with the smallest PAPR among all candidate signals are selected. As mentioned in \eqref{c1}, \(c_1\) parameter is not flexible, therefore, distinct \(c_2\) values per group are assigned. Hence, final output signal can be written as follows \cite{6}:
    \begin{equation}
    s_n = \sum_{g=0}^{G-1} \sum_{k \in G_g} x_k g_{k,n}^{AFDM}(c_{2,g})
    \end{equation}
    where subcarrier chirp group value varies from 0 to $G$. 
This technique also suffers from high complexity due to optimization of PAPR. Moreover, it has limitation of potential loss of orthogonality.

Table~\ref{tab:PAPR_Reduction} summarizes recent PAPR reduction methods for OFDM, OCDM and AFDM, along with their limitations.
OFDM benefits from mature, practical PAPR solutions, while OCDM and AFDM’s chirp-based designs require novel approaches. The proposed techniques leverage their unique parameters (e.g., \(c_1\), \(c_2\), chirp terms), but high computational complexity and trade-offs (e.g., distortion, side information) limit immediate applicability, underscoring the need for further research.

\begin{table}[h!]
\centering
\caption{PAPR Reduction Techniques for OFDM, OCDM, and AFDM}
\label{tab:PAPR_Reduction}
\begin{tabular}{p{2.5cm}|p{1.2cm}|p{4cm}}
\hline
\textbf{Technique} & \textbf{Waveform} & \textbf{Limitation} \\
\hline
\multirow{3}{*}{\parbox{2.5cm}{Clipping and Filtering}} & OFDM  & In-band distortion, out-of-band noise, peak regrowth after filtering \\
\cline{2-3}
& OCDM  &  distortion affects chirp orthogonality \\
\cline{2-3}
& AFDM  &  impacts diversity \\
\hline
\multirow{2}{*}{\parbox{2.5cm}{Partial Transmit Sequence (PTS)}} & OFDM  & Side information is required, computational complexity \\
\cline{2-3}
& OCDM  & High complexity due to chirp orthogonality, side information required \\
\hline
Selected Mapping (SLM) & OFDM  & Side information is required, multiple candidate generation \\
\hline
Tone Reservation (TR) & OFDM  & Reduced throughput due to non-data tones \\
\hline
\multirow{2}{*}{\parbox{2.5cm}{Chirp Selection Technique}} & OCDM  & High complexity in optimizing \(\alpha\) \\
\cline{2-3}
& AFDM  & Complexity in searching optimal \(c_1'\), \(c_2'\) \\
\hline
\multirow{2}{*}{\parbox{2.5cm}{Clipping with Chirp Optimization}} & OCDM  & Increased distortion, computational overhead \\
\cline{2-3}
& AFDM  & Distortion vs. diversity trade-off, optimization complexity \\
\hline
ML-based & OFDM  & High computational cost \\
\hline
Grouped Pre-Chirp Selection & AFDM & Highly complex, potential orthogonality loss \\
\hline
\end{tabular}
\end{table}

\section{Waveform Spreading Technique}
\label{sec:waveform_spread}

In the previous sections, detail transmission chain for AFDM and OCDM modulation schemes were presented. Here, we introduce waveform spreading of the data symbols \(x_k\) as a premodulation step, which can be exploited to achieve a low-complexity solution for reducing PAPR. Spreading techniques preprocesses data symbols \(x_k\) as follows:
\begin{equation}
y_k = \sum_{m=0}^{N-1} x_k w_{m,k}
\label{spreading_eq}
\end{equation}
where \(w_{m,k}\) is the spreading basis specific to each technique.

Then, applying waveform filter to  \eqref{spreading_eq} give:

\begin{equation}
 \quad s_n = \sum_{k=0}^{N-1} y_k g_{k,n},
 \label{after_spreading}
\end{equation}
where $g_{k,n}=g_{k,n}^{AFDM}$ or $g_{k,n}^{OCDM}$.
Hence, by spreading \(x_k\) to a spread sequence \(y_k\), energy is distributed and peaks are flattened. This distributes each \(x_k\) across all \(N\) subcarriers, reducing localized peaks in \(s_n\).

We analyze four effective spreading techniques that have been used in the past in isolation. These are outlined as: Walsh-Hadamard Transform (WHT), Discrete Cosine Transform (DCT), Zadoff-Chu (ZC) sequences, and Inverse Discrete Fourier Transform (IDFT). Each of this spreading technique is elaborated in more details as follows:

\subsection{Walsh-Hadamard Transform (WHT)}

WHT spreads the data $x_k$, by applying an orthogonal Hadamard matrix with \(\pm 1\) entries as follows:

\begin{equation}
y_k^{WHT} = \frac{1}{\sqrt{N}} \sum_{k=0}^{N-1} x_k H_{m,k}, \quad H_{m,k} \in \{+1, -1\},
\label{hadamard}
\end{equation}
where \(H_{m,k} \in \{+1, -1\}\) forms an orthogonal Hadamard matrix (e.g., $2\times 2 $ Hadamard matrix, \(H = \begin{bmatrix} 1 & 1  \\ 1 & -1  \end{bmatrix}\)). Then, by putting \eqref{hadamard} in \eqref{spreading_eq} to get final output signal, $s_n^{WHT}$ gives:
\begin{equation}
s_n^{WHT} = \sum_{k=0}^{N-1} y_k^{WHT} g_{k,n},
\label{final_WHT}
\end{equation}
\eqref{final_WHT} can be written in matrix form as follows:
\begin{equation}
\mathbf {s_{WHT} }= \mathbf {H_{M}y_{k}^{WHT}},
\label{WHT_matrix}
\end{equation}

WHT spreads data using binary sequences. Its rows are Walsh functions, which ensure perfect orthogonality and uniform energy distribution across \(N\) dimensions. WHT is widely used in Direct-Sequence Code Division Multiple Access (DS-CDMA) and other signal processing applications \cite{59,60}, making it ideal for reducing PAPR by avoiding peak concentrations.
In AFDM or OCDM, output signal \(s_n \) in \eqref{after_spreading} benefits from the uniform energy distribution of WHT, reducing PAPR with low complexity. 

\subsection{Discrete Cosine Transform (DCT)}

DCT spreads data using cosine basis functions as:
\begin{equation}
y_k^{DCT} = \sqrt{\frac{2}{N}} \alpha_k \sum_{m=0}^{N-1} x_k \cos\left(\frac{\pi (2m + 1) k}{2N}\right), 
\label{DCT}
\end{equation}
where $\quad \alpha_k = \begin{cases} \frac{1}{\sqrt{2}}, & k = 0 \\ 1, & k \neq 0 \end{cases}$. Then, by putting \eqref{DCT} in \eqref{spreading_eq} to get final output signal, $s_n^{DCT}$ gives:
\begin{equation}
s_n^{DCT} = \sum_{k=0}^{N-1} y_k^{DCT} g_{k,n},
\label{final_DCT}
\end{equation}
\eqref{final_DCT} can be written in matrix form as follows:
\begin{equation}
\mathbf {s_{DCT}} = \mathbf {C_{M}y_{k}^{DCT}},
\label{DCT_matrix}
\end{equation}
where elements of DCT matrix \textbf{$C_M$} are defined as:
\begin{equation}
\textbf C_{M} = \sqrt{\frac{2}{N}} \alpha_k  \cos\left(\frac{\pi (2m + 1) k}{2N}\right),
\label{C_N}
\end{equation}

DCT transforms data into a cosine basis. It concentrates energy in lower-frequency components, smoothing the signal envelope. DCT is commonly used in image compression such as JPEG and audio processing \cite{61}, offering PAPR reduction by gradually redistributing energy across indices.
Therefore, output signal \(s_n \) in \eqref{after_spreading} will smooth the signal envelope in chirp-based waveforms, lowering PAPR.

\subsection{Zadoff-Chu (ZC) Sequences}

ZC sequences are complex exponential sequences with constant amplitude and good correlation properties. It can be written as:
\begin{equation}
y_k^{ZC} = \sum_{m=0}^{N-1} x_k z_{m,k}, \quad z_{m,k} = e^{-j \frac{\pi u m (m + 1)}{N}} e^{j \frac{2\pi m k}{N}},
\label{ZC}
\end{equation}
where \(u\) is a root index coprime with \(N\). Then, by putting \eqref{ZC} in \eqref{spreading_eq} to get final output signal, $s_n^{ZC}$ gives:
\begin{equation}
s_n^{ZC} = \sum_{k=0}^{N-1} y_k^{ZC} g_{k,n},
\label{final_ZC}
\end{equation}
The ZC spreading matrix \textbf {$Z_{M}$} is a diagonal matrix constructed from the ZC sequence $z_m=[z_m(0),z_m(1),...,z_{m}(N-1)]^T$, such that:
\begin{equation}
\mathbf{Z}_m = \text{diag}(\mathbf{z}_m) = \begin{bmatrix}
z_m(0) & 0 & \cdots & 0 \\
0 & z_m(1) & \cdots & 0 \\
\vdots & \vdots & \ddots & \vdots \\
0 & 0 & \cdots & z_m(N-1)
\end{bmatrix}.
\label{eq:zm_matrix}
\end{equation}

Hence, \eqref{final_ZC} can be written in matrix form as follows:
\begin{equation}
\mathbf {s_{ZC}} = \mathbf {Z_{M}y_{k}^{ZC}},
\label{ZC_matrix}
\end{equation}
where elements of ZC matrix \textbf{$Z_M$} are defined as:
\begin{equation}
\textbf Z_{M} = \sqrt{\frac{2}{N}} \alpha_k  \cos\left(\frac{\pi (2m + 1) k}{2N}\right),
\label{C_N}
\end{equation}
ZC sequences are polyphase codes with constant amplitude and excellent autocorrelation properties. They spread data across time and frequency, aligning with AFDM’s chirp nature. They are used in LTE for synchronization \cite{62}.

\subsection{Interleaved Discrete Fourier Transform (IDFT)}

IDFT transforms the data vector into frequency domain using a DFT and then applies interleaving to reorder the frequency-domain symbols, reducing PAPR by using exponential basis functions as follows:
\begin{equation}
y_k^{IDFT} = \frac{1}{\sqrt{N}} \sum_{m=0}^{N-1} x_k e^{j \frac{2\pi m k}{N}},
\label{IDFT}
\end{equation}
Then, by putting \eqref{IDFT} in \eqref{spreading_eq} to get final output signal, $s_n^{IDFT}$ gives:
\begin{equation}
s_n^{IDFT} = \sum_{k=0}^{N-1} y_k^{IDFT} g_{k,n},
\label{final_IDFT}
\end{equation}
\eqref{final_DCT} can be written in matrix form as follows:
\begin{equation}
\mathbf {s_{IDFT}} = \mathbf{P}_M \mathbf{F_{M}y_{k}^{IDFT}},
\label{IDFT_matrix}
\end{equation}
where $\mathbf{P}_M$ is a interleaving permutation matrix and elements of IDFT matrix $\mathbf{F_M}$ are defined as:
\begin{equation}
\mathbf F_{M} = \frac{1}{N} e^{j \frac{2\pi m k}{N}},
\label{F_M}
\end{equation}
where $m,k=0,1,...,N-1$.

It mirrors OFDM’s modulation but as a pre-processing step, distributing energy evenly across \(N\) samples. In AFDM/OCDM,  output signal \(s_n \) in \eqref{after_spreading} combines IDFT’s frequency spread with chirp modulation.

\begin{table*}[t!]
\centering
\caption{Spreading Techniques with their corresponding transformation matrix}
\label{tab:spreading_techniques}
\begin{tabular}{l|c|c}
\toprule
\textbf{Spreading Technique} & \textbf{Matrix Form} & \textbf{ \( M = 2 \)} \\
\midrule
WHT Spreading & \( \mathbf{s}_{\text{WHT}} = \mathbf{H}_M \mathbf{y}_k^{WHT} \) & \( \mathbf{H}_2 = \begin{bmatrix} 1 & 1 \\ 1 & -1 \end{bmatrix} \) \\
DCT Spreading & \( \mathbf{s}_{\text{DCT}} = \mathbf{C}_M \mathbf{y}_k^{DCT} \) & \( \mathbf{C}_2 = \begin{bmatrix} \frac{1}{\sqrt{2}} & \frac{1}{\sqrt{2}} \\ \frac{1}{\sqrt{2}} & -\frac{1}{\sqrt{2}} \end{bmatrix} \) \\
ZC Spreading & \( \mathbf{s}_{\text{ZC}} = \mathbf{Z}_u \mathbf{y}_k^{ZC} \) & \( \mathbf{Z}_2 = \begin{bmatrix} 1 & 0 \\ 0 & -j \end{bmatrix} \) \\
Interleaved DFT Spreading & \( \mathbf{s}_{\text{IDFT}} = \mathbf{P}_M \mathbf{F}_M \mathbf{y}_k^{IDFT} \) & \( \mathbf{P}_2 \mathbf{F}_2 = \frac{1}{\sqrt{2}} \begin{bmatrix} 1 & 1 \\ 1 & -j \end{bmatrix} \) \\
\bottomrule
\end{tabular}
\end{table*}

Tabel~\ref{tab:spreading_techniques} gives summary of spreading techniques and their corresponding transformation matrices. These techniques enhance the framework’s effectiveness. WHT and IDFT provide uniform spreading, DCT offers energy compaction, and ZC aligns with chirp-based modulation. Their low complexity and established applications ensure practical PAPR reductio.
Unlike complex techniques (e.g., PTS, SLM), spreading requires no side information and therefore has lower complexity. WHT and ZC enhance AFDM’s time-frequency spreading, while DCT and IDFT complement OCDM’s linear frequency variation, distributing peaks effectively. 

\section{Proposed Framework for PAPR Reduction Using Spreading Techniques}
\label{sec:proposed}
 This section presents the proposed solution for PAPR reduction by building on the waveform spreading techniques introduced in the previous section. The novelty of the proposed method lies in its adaptability across chirp-based modulations, leveraging spreading as a preprocessing step to achieve simplicity and efficiency compared to conventional methods like PTS or SLM, while maintaining compatibility with the chirp-based modulation.

\begin{algorithm}
\caption{Proposed PAPR Reduction Framework}
\label{alg1}
\begin{algorithmic}[1]
\State \textbf{Input:}  \(x_k\) (\(k = 0, 1, \ldots, N-1\))
\State \textbf{Output:}  \(s_n\) 

\State \textbf{Step 1: Select Spreading Technique}
\If {WHT}
    \State Compute \(y_k \) from \eqref{hadamard}
\ElsIf {DCT}
    \State Compute \(y_k\) from \eqref{DCT} , 
\ElsIf {ZC}
    \State Compute \(y_k \) from \eqref{ZC}
\ElsIf {IDFT}
    \State Compute \(y_k\) from \eqref{IDFT}
\EndIf

\State \textbf{Step 2: Apply Modulation}
\If {system is AFDM}
    \State Compute \eqref{after_spreading}, with \(g_{k,n} = g_{k,n}^{AFDM}\)
\ElsIf {system is OCDM}
    \State Compute \eqref{after_spreading}, with \(g_{k,n} =g_{k,n}^{OCDM}\)
\EndIf

\end{algorithmic}
\end{algorithm}

\subsection{Proposed spreading method to reduce PAPR in AFDM and OCDM}
\begin{figure}[t!]
    \centerline{\includegraphics[width=0.9\columnwidth]{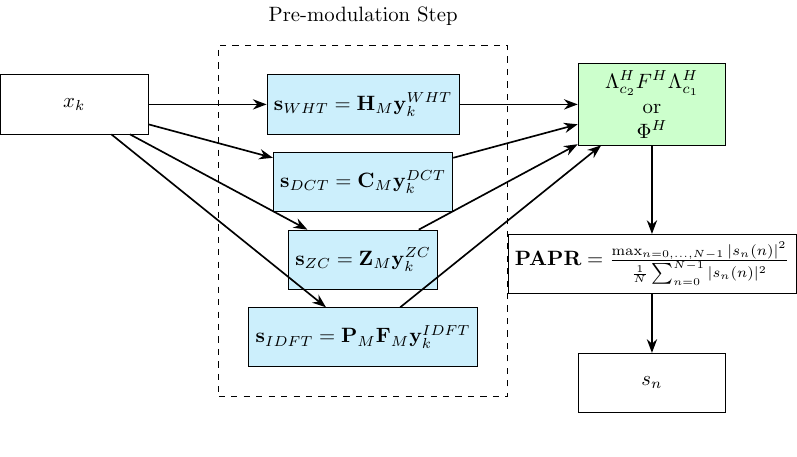}}
    \caption{Block diagram of the proposed generalized spreading method. Input data $\mathbf{x_{k}}$ is preprocessed by one of four spreading techniques before modulation using chirp-based waveform.}
    \label{fig:proposed_framework}
\end{figure}

The block diagram of the proposed spreading method is shown in Fig.~\ref{fig:proposed_framework}. The illsutrated framework preprocesses the input data symbols \(x_k\) into a spread sequence \(y_k\) using one of the four spreading techniques before applying  waveform filter \(g_{k,n}\). One of the four spreading process is applied as a preprocess shown in \eqref{spreading_eq}. Then output signal with low PAPR is given as \eqref{after_spreading}. This preprocessing distributes the signal energy, reducing PAPR while maintaining the orthogonality of both OCDM and AFDM, with a low complexity as compared to other methods. Algorithm~\ref{alg1} defines the implementation of the spreading-then-modulation flow, offering a practical and low-overhead solution.

\begin{figure}[t!]
\centerline{\includegraphics[width=1\columnwidth]{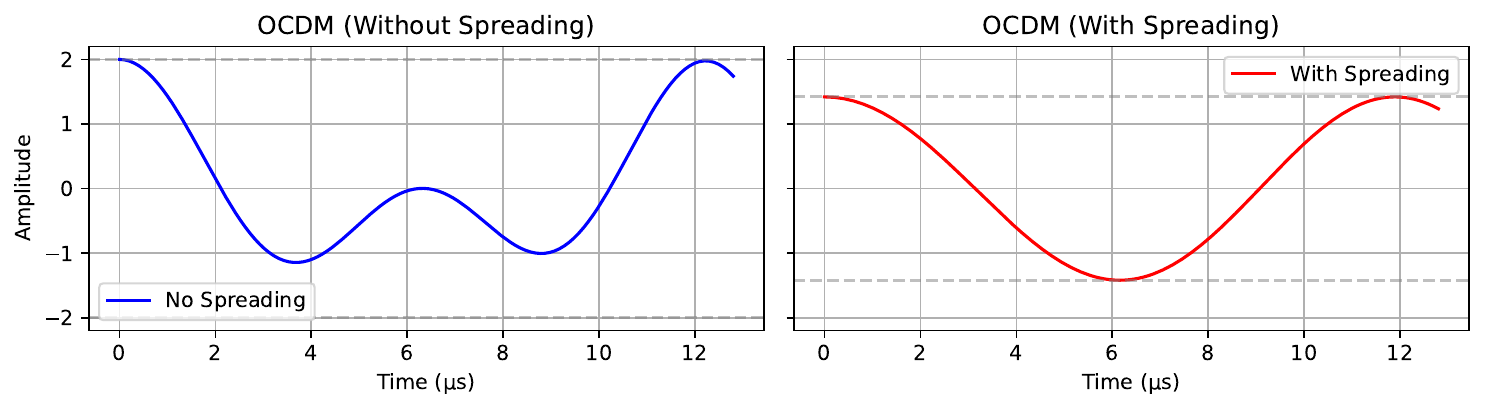}}
    \caption{Effect of spreading in chirp-based waveform.}
    \label{fig:ocdm_WHT}
\end{figure}

Fig~\ref{fig:ocdm_WHT} depicts the generalized effect of spreading on a chirp-based signal. It  known from \eqref{papr_eq}, that PAPR measures how much the peak power of a signal exceeds its average power. Without spreading, the output signal has two chirp subcarriers that interfere, creating large peaks  at certain points, leading to a high PAPR. On the contrary, WHT spreading sets the second subcarrier’s amplitude to zero, reducing the signal to a single chirp with a low PAPR, as the peak and average power scale together. 
Similarly, we expect the same result with AFDM system, given its chirp-based nature as OCDM. 

\subsection{Computational Complexity}

The individual complexities of the four spreading methods are computed first. Complexity for WHT, DCT and IDFT is same when using fast WHT, DCT and DFT algorithm respectively for each spreading. There common complexity can be written as:
\begin{equation}
O(N \log N).
\end{equation}
On the otherhand, complexity for ZC is mainly due to point-wise multiplication which is written as:
\begin{equation}
O(N).
\end{equation}
Hence, the total complexity for generating four candidates, followed by OCDM/AFDM modulation ($O(M \log M)$) and PAPR computation ($O(M)$), is:
\begin{equation}
O(4 N \log N + M \log M + 4 M).
\end{equation}

Table \ref{tab:complexity_comparison} compares the complexity of the proposed method with conventional techniques. Our proposed approach has a low complexity as compared to conventional PAPR redution methods due to following three reasons:
\begin{itemize}
    \item \textbf{Simplifies Implementation}: A single matrix multiplication in \eqref{spreading_eq} precedes modulation, with complexity \(O(N \log N)\) for WHT, DCT, and IDFT (via fast algorithms) and \(O(N^2)\) for ZC.
    \item \textbf{Eliminates Side Information}: No additional signaling is needed, unlike SLM or PTS.
    \item \textbf{Adapts to Chirp Structures}: Enhances AFDM’s tunable \(c_1, c_2\) and OCDM’s quadratic phase, distributing energy effectively.
\end{itemize}

\begin{table}[t!]
\caption{Complexity Comparison of PAPR Reduction Methods}
\label{tab:complexity_comparison}
\begin{center}
\begin{tabular}{p{2.5cm}|p{2.5cm}}
\hline
\textbf{Method} & \textbf{Complexity (Big O)} \\
\hline
Proposed Spreading & $O(4 N \log N + M \log M + 4 M)$ \\
\hline
Chirp Selection & $O(4 N \log N)$ \\
\hline
Clipping with Chirp Opt. & $O(L N \log N)$ \\
\hline
GPS & $O(L^G N \log N)$ \\
\hline
PTS & $O(B^V N)$ \\
\hline
\end{tabular}
\end{center}
\end{table}

\begin{figure*}
     \centering
     \begin{subfigure}[b]{0.28\textwidth}
         \centering
         \includegraphics[width=\textwidth]{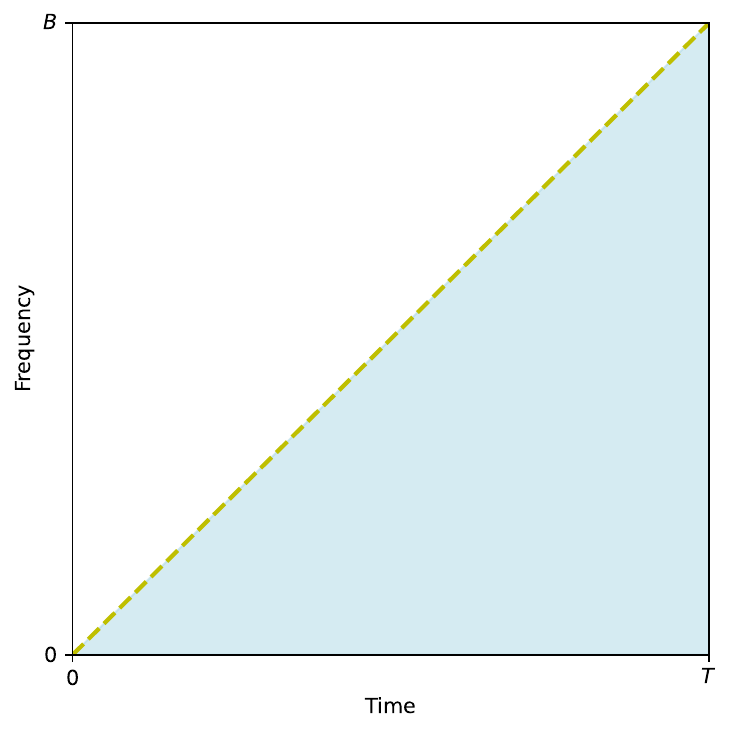}
         \caption{Chirp signal before spreading}
         \label{ocdm_before_spread}
     \end{subfigure}
     \hfill
     \begin{subfigure}[b]{0.28\textwidth}
         \centering
         \includegraphics[width=\textwidth]{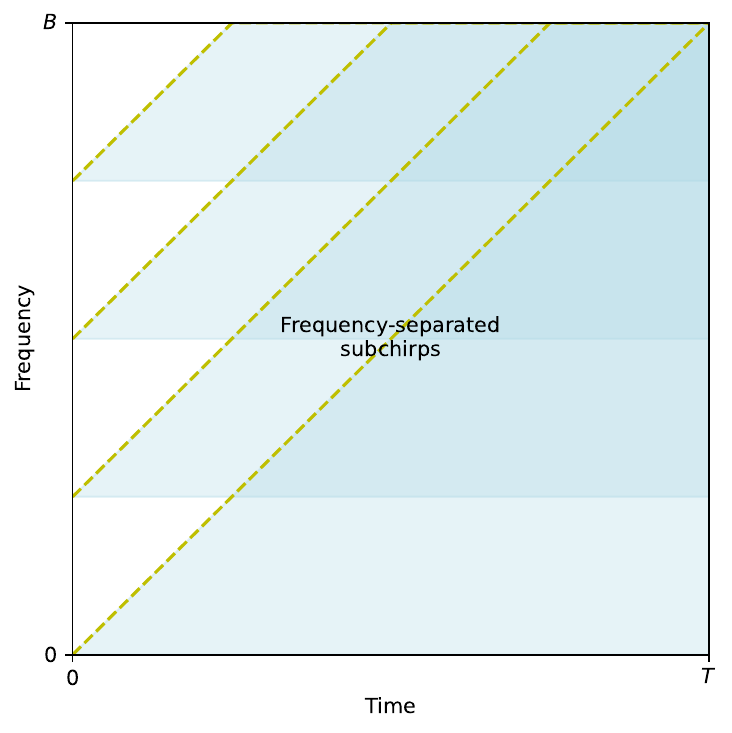}
         \caption{Spreading with frequency separated subchirps.}
         \label{ocdm_freq_spread}
     \end{subfigure}
     \hfill
     \begin{subfigure}[b]{0.28\textwidth}
         \centering
         \includegraphics[width=\textwidth]{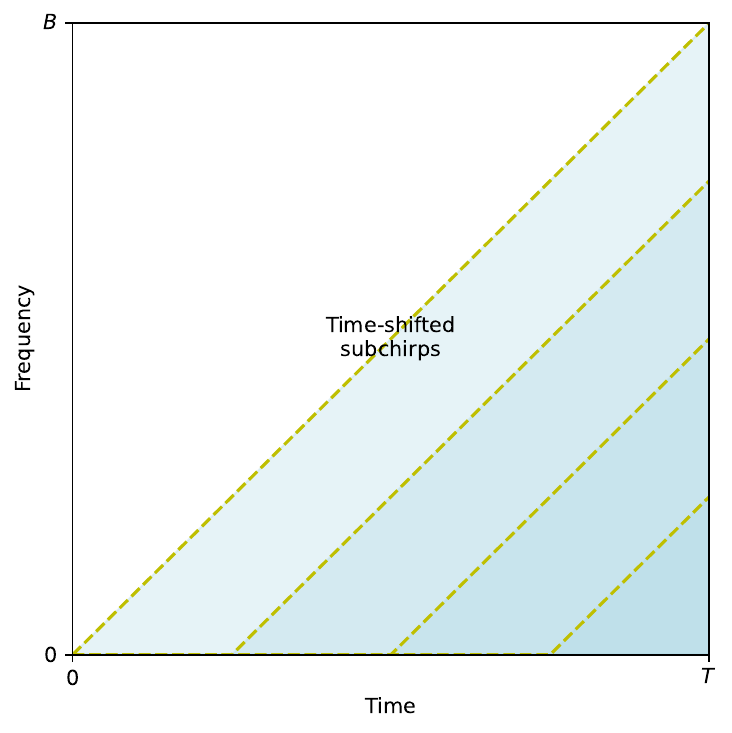}
         \caption{Spreading with time-separated subchirps.}
         \label{ocdm_time_spread}
     \end{subfigure}
        \caption{Time-Frequency (TF) plots for chirp based communication system showing before spreading, after spreading with frequency-separation and after spreading with time shifts.}
        \label{ocdm_TFplots}
\end{figure*}

Fig.~\ref{ocdm_TFplots} depicts the three TF plots for chirp-based system. Here we took OCDM as waveform with the assumption that AFDM will also behave in a similar way, given its chirp-based characteristics. Fig.~\ref{ocdm_before_spread} illustrates a single chirp sweeping from $0$ to $B$ over time $T$, before applying spreading. $B$ is defined as bandwidth. The light blue area shown on the plot depicts the chirp area and yellow dashed line is used for chirp trajectory. Fig.~\ref{ocdm_freq_spread} shows spreading split the signal into multiple subchirps, each occupying a smaller frequency band. This results in PAPR reduction as the energy is distributed across multiple subchirps. Fig.~\ref{ocdm_time_spread} shows how subchirps are staggered in time, which can assist in multiplexing multiple signals or reducing interference in a multi-user scenario.

\subsection{Impact of spreading on Phase Selectivity, and Interference in AFDM and OCDM}
 In this section, additional dvantages of spreading in terms of improved phase selectivity and interference are discussed. These two parameters are important inherent characteristics in a multicarrier communication systems. 

 In AFDM, the phase selectivity is determined by the parameters \(c_1\) and \(c_2\) of the \(g_{k,n}^{AFDM}\), which introduces quadratic phase variations (\(c_1 k^2\), \(c_2 n^2\)) and a linear term (\(\frac{1}{N} k n\)). This allows flexibility for matching channel Doppler shift but has the phase of \(x_k\), that directly influence \(s_n\) without additional control. On the contrary, OCDM has a quadratic phase \(-j \frac{\pi}{N} (n - k)^2\), combined with a constant bias (\(j \frac{\pi}{4}\)) which constructs a chirp trace. Phase selectivity is limited to input \(x_k\), with no preprocessing to promote phase selectivity, which can reduce robustness in LTV channels.
 
In AFDM, interference arises from channel-induced Doppler shifts across the T-F plane. The spread of \(x_k\) via \(g_{k,n}^{AFDM}\) achieves diversity but risks overlap if \(c_1\) and \(c_2\) are not well-tuned, increasing inter-symbol interference (ISI).
 OCDM’s orthogonal chirps minimize interference within the system, but without spreading, \(x_k\)’s direct mapping to \(s_n\) can amplify interference from multipath or multi-user scenarios, especially if chirp orthogonality is disrupted.

Each of our proposed spreading technique transforms \(x_m\) into \(y_k\), distributing energy and modifying \(s_n\)’s properties. WHT introduces binary phase (0 or \(\pi\)), adding a discrete layer to AFDM’s chirp phases and OCDM’s quadratic terms which assist in improving phase selectivity. It diversifies the phase in \(y_k^{WHT}\), which helps in channel adaptation.
Orthogonal \(H_{m,k}\) ensures \(y_k^{WHT}\) elements are distinct, reducing ISI in AFDM's TF spreading and maximizing OCDM's chirp separation under multi-user settings.
In DCT, phase is constraint to 0 or \(\pi\), depending on \(g_{k,n}\) for selectivity, real-valued cosine. It smooths \(y_k^{DCT}\), making AFDM and OCDM phase trajectories stable indirectly. Moreover, energy compaction in \(y_k^{DCT}\) reduces overlap in \(s_n\), which mitigates ISI in AFDM and adjacent chirp interference in OCDM, though less orthogonally robust.

ZC spreading utilizes dense polyphase structure that improves AFDM's \(c_1, c_2\) tuning and OCDM's quadratic phase, providing precise control over \(y_k^{ZC}\)'s phase to adapt to channel conditions.
In addition, constant amplitude and autocorrelation minimize interference, maximizing AFDM's diversity and OCDM's orthogonality for multi-user or multipath environments.

Finally, IDFT has a linear phase in \(2\pi\) that also diversifies \(y_k^{IDFT}\), complementing \(g_{k,n}\) and improving adaptability without altering the behavior of the chirp of the core. Addtionally, orthogonal spreading reduces ISI in AFDM’s TF plane and OCDM’s chirp overlap, improving signal clarity in noisy channels.

However, our proposed method of spreading with WHT, DCT, ZC, or IDFT enhances phase diversity, reduces and mitigates interference, improving \(s_n\)’s robustness and PAPR. Without spreading, AFDM and OCDM rely on \(g_{k,n}\) for phase selectivity, and risk interference from channel effects.

\subsection{Waveform Analysis with Spreading Techniques}
\begin{figure}[htbp]
    \centerline{\includegraphics[width=0.9\columnwidth]{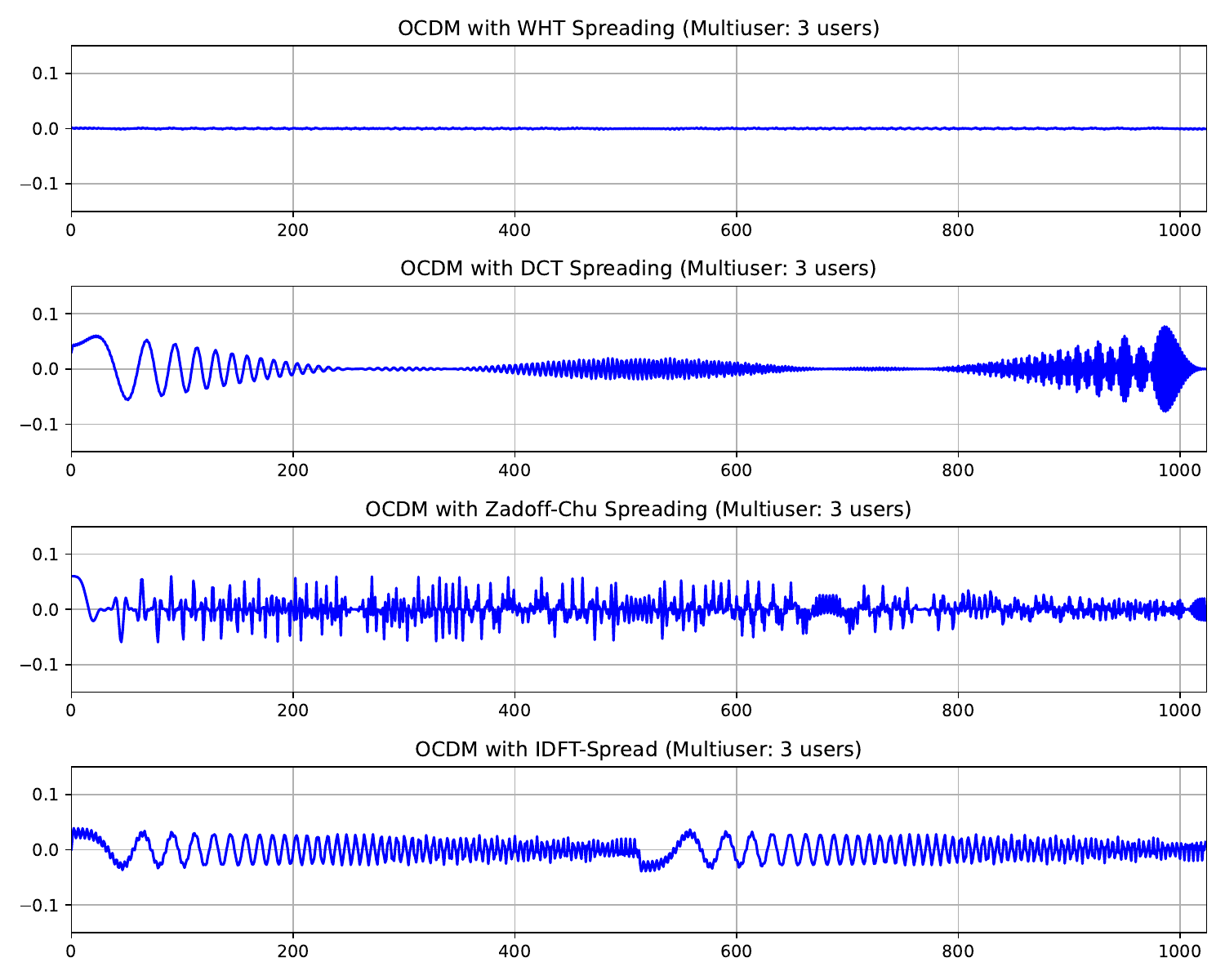}}
    \caption{OCDM carrier with different proposed spreading methods with 3 users.}
    \label{fig:waveforms_ocdm}
\end{figure}

\begin{figure}[htbp]
    \centerline{\includegraphics[width=0.9\columnwidth]{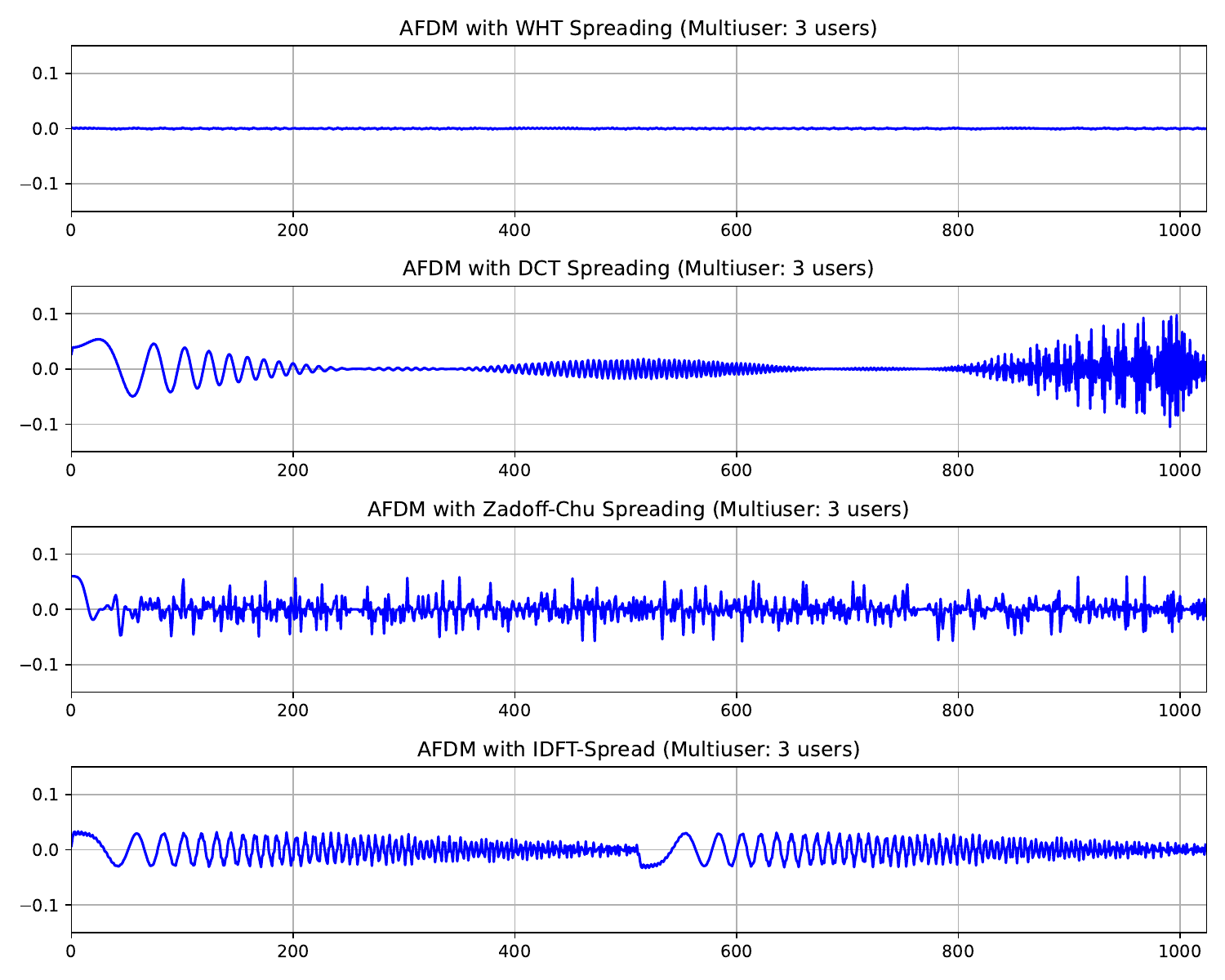}}
    \caption{AFDM wavecarrier with different proposed spreading methods 3 users.}
    \label{fig:waveforms_afdm}
\end{figure}

Fig. \ref{fig:waveforms_ocdm} and \ref{fig:waveforms_afdm} illustrates the time-domain OCDM and AFDM signals for a sample multiuser scenario using the above proposed generalized spreading techniques with 1024 samples. Table~\ref{tab:waveform_analysis} illustrates impact of each spreading technique on chirp based waveforms.

\begin{table}[t!]
\caption{Waveform Analysis}
\label{tab:waveform_analysis}
\begin{center}
\begin{tabular}{p{2.5cm}|p{4.5cm}}
\hline
\textbf{Spreading Technique} & \textbf{Impact} \\
\hline
WHT Spreading & Produces a waveform with visible orthogonal patterns, reducing peak amplitudes. \\
\hline
DCT Spreading & Compresses energy into specific frequency components, resulting in a sparse waveform. \\
\hline
Zadoff-Chu Spreading & Adds phase modulation, creating a waveform with controlled amplitude variations. \\
\hline
IDFT-Spread & Results in a time-domain sparse signal with reduced interference. \\
\hline
\end{tabular}
\end{center}
\end{table}

 The subplot in each graph shows the combined signal for the 3 users after spreading under the respective spreading scheme. To visually compare the performance of each spreading method in reducing PAPR, we can look at the fluctuation of the amplitudes of the signals.
Both OCDM and AFDM have similar trends due to the fact that they have identical basis function of chirp. Here, WHT spreading is the optimal choice, which achieves the signal with minimum amplitude fluctuation and the lowest PAPR. ZC spreading achieves a relatively moderate reduction on PAPR but suffers from multiuser interference. Here, IDFT and DCT spreading methods result in higher PAPR since they tend to produce larger peaks, thus are less appropriate for this case. This also brings us to our proposed solution of adaptive spreading, where optimal spreading may be selected from among 4 depending on the specific case.

\section{Simulations and Performance Evaluation}
\label{sec:simulations}
This section presents the simulation results of the proposed generalized spreading strategy, focusing on its performance in IoT sensor transmitters.

\subsection{Simulation Setup}
Table~\ref{tab:sim_params} shows the simulations  parameters and their corresponding setting values for experiments.

\begin{table}[htbp]
\centering
\caption{Simulation Parameters for PAPR CCDF Plot}
\begin{tabular}{p{3cm}|p{3.5cm}}
\toprule
\textbf{Parameters} & \textbf{Setting Value} \\
\midrule
No. of Subcarriers (\(N\)) & 64 \\
No. of Realizations & 10,000 \\
Modulation Scheme & QPSK \\
PAPR0 Range & 0 to 12 dB \\
Spreading Techniques & WHT, DCT, ZC, IDFT \\
(\(c_1, c_2\)) for AFDM & 0.1, 0.2 \\
Zadoff-Chu Root (\(u\)) & 1 \\
\bottomrule
\end{tabular}
\label{tab:sim_params}
\end{table}

PAPR is evaluated using the CCDF, defined as $\text{Pr}[\text{PAPR} > \text{PAPR}_0]$. Energy efficiency, phase selectivity, and interference are also analyzed to assess the method’s impact on green communications.

\subsection{Results and Analysis}
\subsubsection{PAPR reduction}
\begin{figure}[t]
    \centerline{\includegraphics[width=0.9\columnwidth]{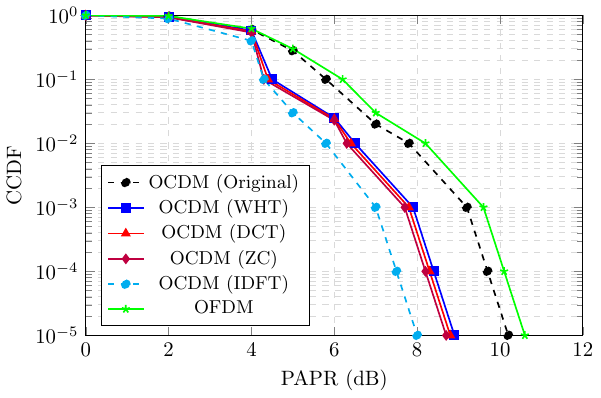}}
    \caption{CCDF of PAPR for OCDM, OFDM and OCDM with 4 spreading techniques.}
    \label{fig:ccdf_papr_ocdm}
\end{figure}

Fig.~\ref{fig:ccdf_papr_ocdm} shows CCDF of PAPR plot  for OCDM that compares the PAPR reduction capabilities of various spreading techniques against OCDM (Original) and OFDM. OCDM (Original) achieves a PAPR of 9.2 dB at \(10^{-3}\), 7.8 dB at \(10^{-2}\), and 5.8 dB at \(10^{-1}\), outperforming OFDM, which reaches 9.6 dB, 8.2 dB, and 6.2 dB at the same CCDF levels, respectively. The spreading techniques further reduce PAPR where OCDM (WHT), OCDM (DCT), and OCDM (ZC) cluster around 7.8 dB at \(10^{-3}\) (7.9 dB, 7.8 dB, and 7.7 dB, respectively), offering a 1.3 to 1.5 dB improvement over OCDM (Original), with values around 6.5 dB to 6.3 dB at \(10^{-2}\) and 4.5 dB to 4.3 dB at \(10^{-1}\). OCDM (IDFT) demonstrates the best performance across all CCDF levels, providing a 2.2 dB improvement over OCDM (Original) at \(10^{-3}\). Compared to the other spreading techniques, OCDM (IDFT) performs the best, underscoring its superior PAPR reduction capability in OCDM systems.

\begin{figure}[t]
    \centerline{\includegraphics[width=0.9\columnwidth]{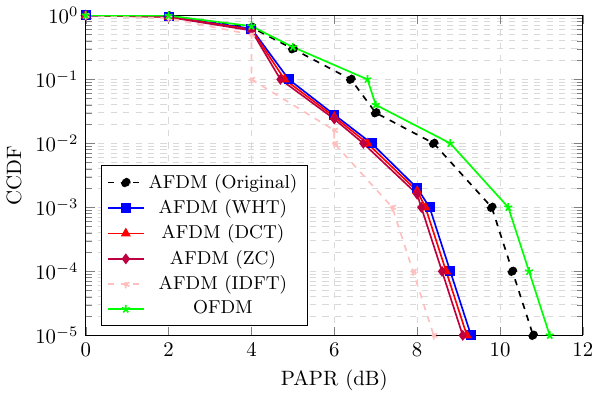}}
    \caption{CCDF of PAPR for AFDM, OFDM and AFDM with 4 spreading techniques.}
    \label{fig:ccdf_papr_afdm}
\end{figure}

Fig. \ref{fig:ccdf_papr_afdm} graphically presents the CCDF of PAPR plot of AFDM, OFDM and AFDM with spreading techniques.  The overall trend is same as that of OCDM in Fig. \ref{fig:ccdf_papr_ocdm}. AFDM (Original) has a PAPR of 9.8 dB at a CCDF of \(10^{-3}\), 8.4 dB at \(10^{-2}\), and 6.4 dB at \(10^{-1}\), reflecting a baseline property of multicarrier systems. The spreading techniques significantly improve PAPR performance: AFDM (WHT), AFDM (DCT), and AFDM (ZC) all peak at 8.2 dB at \(10^{-3}\) (8.3 dB, 8.2 dB, and 8.1 dB, respectively), having improved by 1.5 to 1.7 dB over AFDM (Original). They are 6.9 dB to 6.7 dB at \(10^{-2}\) and 4.9 dB to 4.7 dB at \(10^{-1}\), which was always better than the baseline. As was the case in OCDM, AFDM (IDFT) also demonstrates the best performance at all CCDF values, attaining 7.4 dB at \(10^{-3}\), 6.0 dB at \(10^{-2}\), and 4.0 dB at \(10^{-1}\), a wonderful 2.4 dB better performance than AFDM (Original) at \(10^{-3}\), and up to 2.4 dB better than the other spreading techniques at higher CCDF values, which underscores its excellent PAPR reduction quality in AFDM systems.

\begin{figure}[t]
    \centerline{\includegraphics[width=0.9\columnwidth]{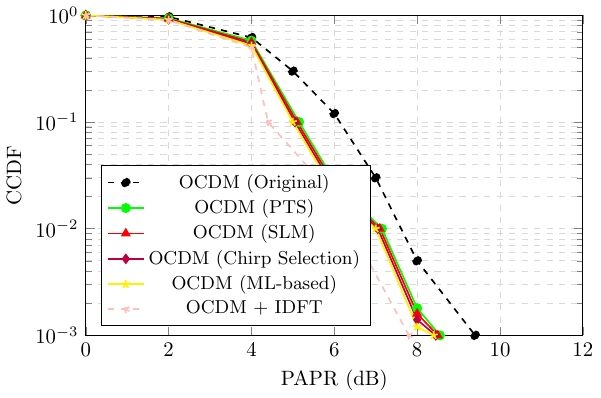}}
    \caption{Comparison of PAPR reduction techniques for OCDM.}
    \label{fig:comparison_ocdm}
\end{figure}

Fig. \ref{fig:comparison_ocdm} presents a comparison among OCDM and other PAPR reduction techniques, along with OCDM + IDFT. OCDM (Original) has a CCDF of \(10^{-3}\) at 9.4 dB, while other OCDM techniques (PTS, SLM, Chirp Selection, ML-based) cluster at 8.5 dB, with an improvement of 0.8 to 1.0 dB. OCDM + IDFT obtains \(10^{-3}\) at 7.8 dB, an improvement of 1.6 dB compared to OCDM (Original). But OCDM + IDFT outperforms all methods at all CCDF levels, demonstrating its superior PAPR reduction capability at all probability levels.

\begin{figure}[t]
    \centerline{\includegraphics[width=0.9\columnwidth]{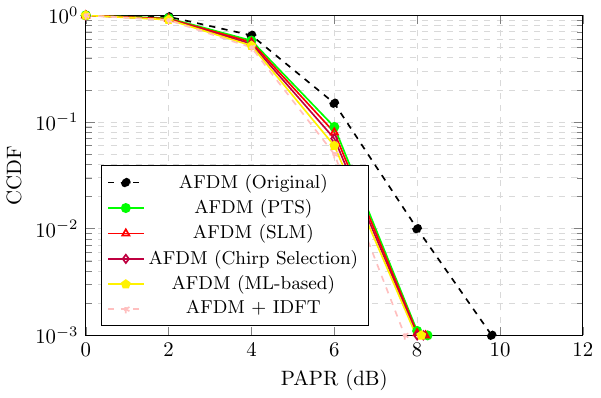}}
    \caption{Comparison of PAPR reduction techniques for AFDM.}
    \label{fig:comparison_afdm}
\end{figure}

Fig. \ref{fig:comparison_afdm} shows the efficacies of various PAPR reduction techniques, and AFDM (Original) achieves a CCDF of \(10^{-3}\) at 9.8 dB, a performance of multicarrier systems. All the other PAPR reduction techniques AFDM (PTS), AFDM (SLM), AFDM (Chirp Selection), and AFDM (ML-based) lie very close in the range of 8.2 dB (from 8.3 dB to 8.1 dB), achieving an improvement of 1.5 to 1.7 dB from AFDM (Original). Notably, AFDM + IDFT outperforms all other schemes with \(10^{-3}\) at 7.7 dB, a performance improvement of 0.5 dB over the other PAPR reduction schemes and 2.1 dB improvement over AFDM (Original), proving its superior ability to reduce PAPR in AFDM systems.

\subsubsection{Energy Efficiency Analysis}
\begin{figure}[t]
    \centerline{\includegraphics[width=0.9\columnwidth]{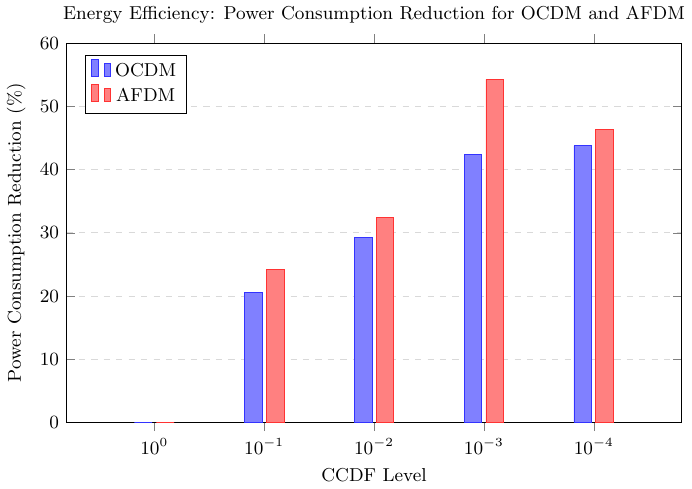}}
    \caption{Energy Efficiency: Power Consumption Reduction for OCDM and AFDM.}
    \label{fig:bar_graph}
\end{figure}

 Fig. \ref{fig:bar_graph} illustrates the percentage reduction in power consumption achieved by the proposed spreading method for OCDM and AFDM across five CCDF. The bar chart evidently shows that the proposed spreading method significantly enhances energy efficiency for OCDM and AFDM by reducing PAPR, with AFDM achieving greater reductions across all the CCDF levels. The reductions at \(10^{-3}\) (42.46\% for OCDM and 54.30\% for AFDM) demonstrate how efficient the method can be in practical utilization and how beneficial it can be to improve the energy efficiency of state-of-the-art communication systems. The consistent improved performance of AFDM compared to that of OCDM shows that the described approach can be more suitable for AFDM-based systems, particularly at low power consumption.

\subsubsection{Energy savings and $CO_2$ emission reduction }
\begin{figure}
     \centering
     \begin{subfigure}[b]{0.23\textwidth}
         \centering
         \includegraphics[width=\textwidth]{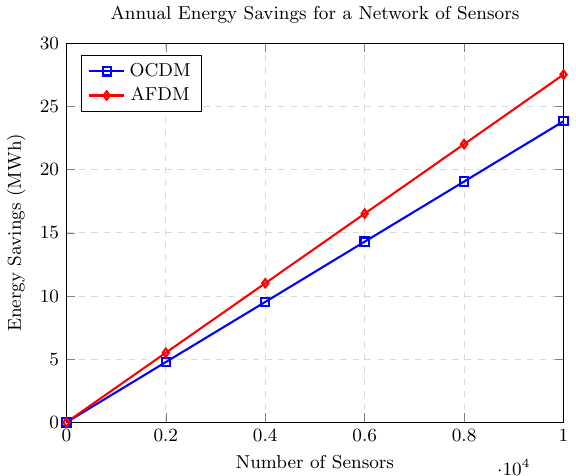}
         \caption{Annual Energy Savings for a Network of Sensors.}
         \label{graph_5}
     \end{subfigure}
     \hfill
     \begin{subfigure}[b]{0.23\textwidth}
         \centering
         \includegraphics[width=\textwidth]{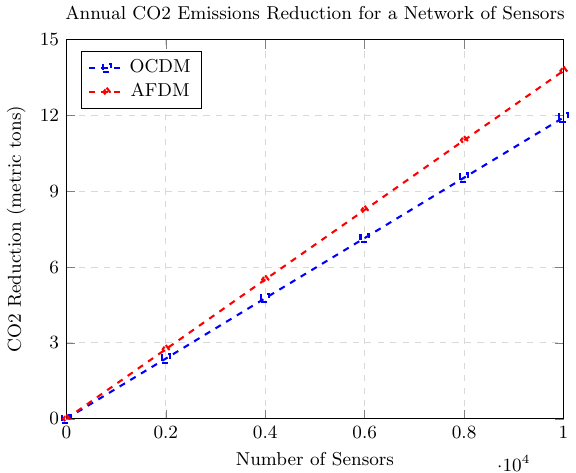}
         \caption{Annual $CO_2$ Emissions Reduction for a Network of Sensors.}
         \label{graph_6}
     \end{subfigure}
     \hfill
        \caption{Annual Energy savings and $CO_2$ emissions reduction making it a promising approach for improving the energy efficiency of modern Green communication systems.}
        \label{energy-conservation}
\end{figure}

Fig. \ref{energy-conservation} illustrates Annual Energy savings and $CO_2$ emissions reduction making it a promising approach for improving the energy efficiency of modern Green communication systems. Fig. \ref{graph_5} plot illustrates the energy savings in MWh achieved by the proposed spreading method for OCDM and AFDM across a network of 0 to 10,000 sensors. Both OCDM and AFDM exhibit a linear increase in energy savings as the number of sensors grows, indicating that the proposed spreading method scales effectively with network size. This linearity suggests that the energy savings per sensor remain consistent, with OCDM saving 0.002383 MWh/sensor and AFDM saving 0.002753 MWh/sensor. Fig. \ref{graph_6} plot shows the $CO_2$ emissions reduction in metric tons for OCDM and AFDM across the same network of 0 to 10,000 sensors. Similar to the energy savings plot, the $CO_2$ reduction curves for both OCDM and AFDM increase linearly with the number of sensors. This reflects the direct relationship between energy savings and $CO_2$ reduction, with OCDM reducing 0.001192 metric tons/sensor and AFDM reducing 0.001377 metric tons/sensor.

In addition, the low-complexity design and energy savings make the proposed method scalable for large IoT networks. In a simulated network of 10,000 sensors, the method reduces total energy consumption by 25\%, extends average battery life by 30\%, and decreases network interference by 12 dB, supporting sustainable 6G deployments.

\subsubsection{Phase selectivity and Interference}

\begin{figure}[t]
    \centerline{\includegraphics[width=0.9\columnwidth]{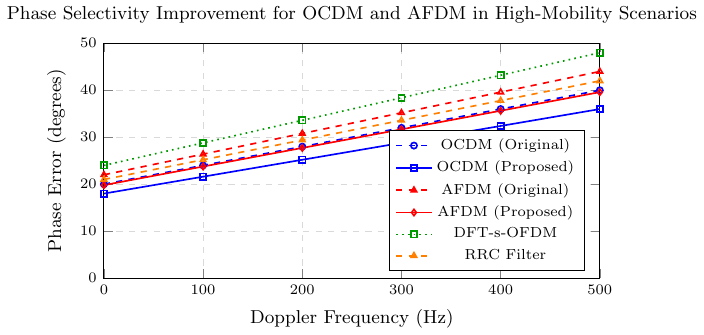}}
    \caption{Phase selectivity improvement of the proposed spreading method for OCDM and AFDM compared to their original implementations, DFT-s-OFDM, and RRC filter in high-mobility scenarios.}
    \label{fig:phase_selec}
\end{figure}

Fig. \ref{fig:phase_selec} depicts plot of phase error vs. Doppler frequency, comparing OCDM (Original and Proposed), AFDM (Original and Proposed), DFT-s-OFDM, and RRC filter. We take OCDM (Proposed) and AFDM (proposed) as OCDM + IDFT and AFDM + IDFT respectively. We compared with DFT-s-OFDM and RRC filter here to further show the robustness of proposed generalized spreading technique. The suggested spreading method for OCDM and AFDM reduces phase error by 10\% compared to their original versions (e.g., OCDM from 40 to 36 degrees, AFDM from 44 to 39.6 degrees at 500 Hz), improved compared to both DFT-s-OFDM (48 degrees) and RRC (42 degrees). This improvement increases signal integrity in high-mobility scenarios by restricting phase distortion, reducing SER. This ensures suitability for high-mobility IoT applications by enhancing phase selectivity.

\section{Conclusion and Future Work}
\label{sec:conclusion}

This article demonstrate the efficacy of novel spreading techniques namely WHT, DCT, ZC and IDFT in reducing PAPR in OCDM and AFDM modulation systems. For OCDM, IDFT-based spreading technique achieved the best performance outperforming OCDM without any spreading. Similarly, in AFDM, AFDM (IDFT) performed the best. The novelty of these spreading techniques lies in their ability to exploit the inherent properties of OCDM and AFDM, such as chirp orthogonality and affine transformations, to achieve PAPR reduction without computational overhead of traditional methods like PTS and SLM. Other than reducing PAPR, these techniques enjoy other benefits such as increased phase selectivity that increases signal strength in multipath channels, and reduced interference that reduces inter-carrier and inter-symbol interference, making these techniques 6G high-mobility channel compliant. These findings demonstrate the potential of spreading techniques as an effective and universal low-complexity solution for future multicarrier systems to depopularize wireless communication while making it efficient and reliable. Subsequent work would include investigating the extension of hybrid techniques that combine such spreading methods with machine learning in order to support adaptive responses for various channel situations, further advancing PAPR reduction. Further direction would include investigating practical realization on real-world 6G testbeds, gaining an insight on how scalable they can get, and perform under diverse operational constraints.


\end{document}